\documentclass[traditabstract]{aa}
\usepackage{graphicx}
\usepackage{natbib}
\usepackage{txfonts}
\usepackage{epsfig}
\usepackage{times}

\newcommand{\amin} {\mbox{$^{\prime}$}}

\def\gsimeq{\hbox{\raise0.5ex\hbox{$>\lower1.06ex\hbox{$\kern-1.07em{\sim}$}$}}} 
\def\lsimeq{\hbox{\raise0.5ex\hbox{$<\lower1.06ex\hbox{$\kern-1.07em{\sim}$}$}}} 

\begin{document}

\title{Galactic outflow driven by the active nucleus and  the origin of the gamma-ray emission in \textsc{NGC 1068}}

\authorrunning{A. Lamastra, et al.}
  \author{A. Lamastra\inst{1}, F. Fiore\inst{1}, D. Guetta\inst{1,2}, L. A. Antonelli\inst{1,3}, S. Colafrancesco\inst{4}, N. Menci\inst{1}, S. Puccetti\inst{1,3}, A. Stamerra\inst{5,6}, L. Zappacosta\inst{1}}
   \offprints{alessandra.lamastra@oa-roma.inaf.it}
   \institute{$^1$ INAF - Osservatorio Astronomico di Roma, via di Frascati 33, 00078 Monte Porzio Catone, Italy\\
   $^2$ Department of Physics and Optical Engineering, ORT Braude College, Karmiel 21982, Israel\\
   	$^3$ ASDC--ASI, Via del Politecnico, 00133 Roma, Italy\\
   	$^4$ School of Physics, University of the Witwatersrand, Private Bag 3, 2050-Johannesburg, South Africa\\
   	$^5$ INAF - Osservatorio Astrofisico di Torino, via Osservatorio, 20, 10025 Pino Torinese, Italy \\
   	 $^6$  Scuola Normale Superiore, Piazza dei Cavalieri 7, 56126 Pisa, Italy \\}
   \date{Received ; Accepted }
   \abstract{We compute the non-thermal emissions produced by relativistic particles accelerated by the AGN-driven shocks in \textsc{NGC 1068}, and we compare the model predictions with the observed $\gamma$-ray and radio spectra . The former is contributed by pion decay, inverse Compton scattering, and bremsstrahlung, while the latter  is produced by synchrotron radiation.
 We derive the $\gamma$-ray and radio emissions by assuming the standard acceleration theory, and we discuss how our results compare with those corresponding to other commonly assumed sources of $\gamma$-ray and radio emissions, like Supernova remnants (SNR) or AGN jets. We find that the AGN-driven shocks observed in the circumnuclear molecular disk of such a galaxy  provide a contribution to the $\gamma$-ray  emission comparable to that provided by the starburst activity when standard particle acceleration efficiencies are assumed, while they can yield the whole $\gamma$-ray emission only when 
the parameters describing the acceleration efficiency and the proton coupling with the molecular gas are tuned to values larger than those assumed in standard, SNR-driven shocks. We discuss the range of 
acceleration efficiencies (for protons and  electrons) and of proton calorimetric fractions required to account for the observed $\gamma$-ray emission in the AGN outflow model. 
\\
   We further compare the neutrino flux expected in our model with constraints from current experiments, and we provide predictions for the detections by the  upcoming KM3NeT neutrino telescope.
This analysis strongly motivates  observations of \textsc{NGC 1068} at $ \gtrsim $ TeV energies with current and future  Cherenkov telescopes in order to gain insight into the nature of the $\gamma$-rays source.

 }

\titlerunning{AGN-driven outflow and the gamma-ray emission in \textsc{NGC 1068}}

 \maketitle
 
\section{Introduction}
\textsc{NGC 1068}  is the brightest,  closest  and best studied Seyfert 2 galaxy. The discovery of its Seyfert 1 nucleus in  polarized light led  \cite{Antonucci85} to propose the AGN unification model. In its central region this source exhibits both starburst and AGN activities. 
Interferometric observations in the millimetre (mm) band identified a circumnuclear disk (CND) $ \sim $200 pc in  radius, surrounded by a $ \sim $2 kpc starburst ring connected to the CND by a bar. 
In X-rays, the  spectrum is  dominated by reflection components of the primary AGN radiation by Compton thick material (i.e. with column density  $N_H>1.5 \times 10^{24}$ cm$^{-2}$), and in particular by a strong K$\alpha$ iron line (EW$ \simeq $1 , \citealt{Matt04}). Recently \cite{Marinucci16} detected a transient flux excess at energies above 20 keV that can be explained by a temporary decrease of $N_H$  along the line of sight. This  event allows to unveil the primary AGN emission and to infer  an intrinsic 2-10 keV luminosity of L$_X=7\times 10^{43}$ erg/s (corresponding to bolometric luminosity L$_{AGN}\simeq 2.1\times 10^{45}$ erg/s, \citealt{Marconi04}). \\
\textsc{NGC 1068}  is a strong $\gamma$-ray emitter. It is the brightest of the few non-blazar galaxies detected by  the {\it Fermi Gamma-ray space telescope}  and with the flattest $\gamma$-ray spectrum \citep{Ackermann12}. Models assuming that the  $\gamma$-ray  emission is entirely due to starburst activity failed to reproduce the observed spectrum \citep{Yoast14, Eichmann15}.
This suggests an alternative/complementary origin for the   $\gamma$-ray emission (see e.g. \citealt{Lenain10}).\\
Interestingly, sub-mm  interferometry of molecular lines in the CND strongly suggests the existence of a giant, AGN-driven outflow which extends to $ \sim $100 pc scale with velocity of  $ \sim $ (100-200) km/s \citep{Krips11, Garcia14}. This outflow can induce shocks in the CND, which, in turn, can accelerate relativistic particles with an efficiency that may exceed that  in  Supernova remnant (SNR)   and could leave observational signatures in different electromagnetic bands \citep{FGQ12,Nims15}.  In addition to primary accelerated electrons, the decay of neutral pions created by  collisions between relativistic protons accelerated by the AGN shocks with ambient  protons may produce a significant 
$\gamma$-ray emission. This hadronic $\gamma$-ray emission  is mostly favored as the dominat  component of the $\gamma$-ray spectrum at energies above $ \simeq $100 MeV. At lower energies  leptonic processes like inverse Compton (IC) scattering  and non-thermal bremsstrahlung can significantly contribute to the $\gamma$-ray  emission.  The same  electrons responsible for IC  and bremsstrahlung emission  spiraling in interstellar  magnetic fields  radiate synchrotron emission in the radio continuum.
The interpretation of the radio emission  as a by-product of the AGN-driven outflow activity could in part explain the deviation of \textsc{NGC 1068} from the  observed correlation between the radio and far infrared (FIR) luminosities of star forming galaxies. The FIR-radio correlation  spans over five orders of magnitude in luminosity, from dwarf galaxies to starburst galaxies \citep{Condon91,Yun01}  and has been explained in terms of FIR emission  related to dust heated by young massive stars, and  radio emission associated to relativistic electrons accelerated in SNR.
NGC1068 is observed to have about four times larger radio luminosity than expected from the radio-FIR correlation \citep{Yun01}.\\
Radio emission from AGN-driven outflows can also explain  the strong correlation  between the kinematics of the ionized gas emission and the radio luminosity observed in obscured radio-quiet quasars at $z\lesssim$1 \citep{Zakamska14}.\\
 In this picture, the $\gamma$-ray and radio luminosities are determined by the energy supplied to relativistic protons and electrons at the shock.
In this paper we examine if the kinetic power of the AGN-driven outflow observed in \textsc{NGC 1068} can account for the observed $\gamma$-ray and radio  luminosities.\\
The paper is organized as follows.  An overview of the observational data, including our {\it Fermi} Large Area Telescope (LAT) data analysis, is given in Section \ref{CND}. 
The physical processes involved in the cosmic-ray (CR) particle energy distributions and the non-thermal  emission  produced by accelerated particles are described in Section \ref{model} . In Section \ref{results} we present our results, discussion and conclusion follow in Section \ref{discussion} and \ref{conclusions}. \\
Throughout the paper, we use a vacuum-dominated cosmological model with $\Omega_m$=0.3, $\Omega_{\lambda}$=0.7, and $H_0$=70 km s$^{-1}$ Mpc$^{-1}$;  and we adopt a distance to \textsc{NGC 1068} of 14.4 Mpc, so that 1$^{\arcsec}\simeq$70 pc.

\section{Observational data}\label{CND}

 In this Section we  review the general structure and physical properties of the central region of \textsc{NGC 1068}, which we will use in the computation of the non-thermal emissions from accelerated particles in the AGN-driven shocks.

\textsc{NGC 1068} has been the target of several observational campaigns over the entire electromagnetic spectrum. \\
In the  radio band the continuum emission is spatially resolved into different structures: a 3 kpc diameter star-forming disk, a kpc-scale radio jet, and a 100 pc-scale jet and compact radio knots \citep[e.g.][]{Gallimore96,Wynn85,Gallimore06,Sajina11,Honig08}.\\
Near-IR  observations of the narrow line region (NLR) revealed that the NLR gas is outflowing in a bicone that follow the  the path of the radio jet \citep{Muller11,Barbosa14}. \\
The distribution and kinematics of the molecular gas in the galaxy disk have been mapped through molecular line surveys \citep{Muller09,Krips11,Garcia14,Barbosa14,Garcia16,Gallimore16}.
 High resolution Atacama Large Millimetre/submillimetre Array (ALMA) observations spatially resolved the kinematics of the molecular gas within the few parsecs from the nucleus (the putative obscuring torus).
The kinematics  of the molecular gas in the torus show strong non-circular motions consistent with a bipolar outflow along the axis of the AGN accretion disk, similar to the outflow inferred for the NLR gas   \citep{Garcia16,Gallimore16}.
At larger scales ALMA observations revealed   a starburst ring of radius  $ \sim $2 kpc and a CND.  In CO maps the CND appears  as an asymmetric ring of  350 pc $\times$ 200 pc size. The ring 
shows a rich substructure  with two prominent knots located east and west of the  position of the AGN.\\
By combining  ALMA and  {\it Herschel} observations \cite{Garcia14} derived the molecular gas mass of the CND from SED fitting of the dust continuum emission. They estimated   M$_{gas} \simeq$(5$\pm$3)$\times$10$^7$M$_{\odot}$, which corresponds to a gas surface density  $\Sigma_{gas}\simeq$(0.01-0.05) g cm$^{-2}$ and to a gas number density  $n_H=$(115-460) cm$^{-3}$  assuming a  cylindrical geometry with radius $R=$350 pc and  vertical scale height $h\simeq$10 pc  \citep{Schinnerer00}. \\
The above estimate of the gas surface density can be used to estimate the  volume averaged interstellar medium (ISM) magnetic field strength of the CND  B$_{ISM}\simeq$(30-110) $\mu$G    from their empirical scaling relation  B$_{ISM}\simeq$(30-110) $\mu$G  \citep{Robishaw08,Mcbride14}. \\

\subsection{Properties and powering source of the molecular outflow}
A sizeable fraction of the total gas content in the CND is involved in a massive outflow \citep{Krips11,Garcia14,Barbosa14}. 
The most relevant outflow properties in the derivation of the non-thermal emissions from shock accelerated particles is the kinetic energy injected into the ISM  during the outflow time-scale (see Sect. \ref{spectra}). The kinetic luminosity  of the outflow is given by its bulk motion, and can be derived from the expression:
 \begin{equation}\label{Lkin}
 L_{kin}=\frac{1}{2}\times\frac{dM_{out}}{dt}\times v_{out}^{2}
 \end{equation} 
where $d{M}_{out}/dt$ is the mass outflow rate and $v_{out}$ is the outflow velocity. \\
From the broad CO(3-2) components in their spectra \cite{Garcia14} estimated the outflowing gas mass  $M_{out}$=1.8$\times$10$^{7}$ M$_{\odot}$,  
 the average radial extent of the outflow  $R_{out}$=100 pc, and the projected radial outflow velocity  $v_{out,p}\simeq$100 km/s. 
Assuming a multi-conical outflow geometry  they derived the mass outflow rate as $d{M}_{out}/dt=3\times v_{out,p}\times M_{out}/R_{out}\times \tan\alpha \simeq$63 M$_{\odot}$ yr$^{-1}$; here $\alpha$ is the unknown angle between the outflow and the line of sight. They assumed that the outflow is coplanar with the main disk, i.e.  $ \tan\alpha  $=$ 1/\tan i $ where $i$=41$^{\circ}$ is the disk inclination angle, and, by replacing $v_{out}$ with $v_{out,p}/\cos{\alpha}$ in  eq. (\ref{Lkin}),   they estimated an outflow kinetic luminosity  equal to  $L_{kin} \simeq $5$\times$10$^{41}$ erg/s.
The main uncertainty of  this estimate is due to the unknown inclination angle between the outflow and the line of sight.\\
Here we  estimate  $L_{kin}$ without any assumptions  about the inclination angle $\alpha$. To this purpose, assuming that the outflow is isotropic we infer  a maximum outflow velocity\footnote{We compute the maximum velocity as in \citealt{Rupke11}} $v_{max}\simeq$200 km/s  directly from  the CO(3-2) spectra obtained with the Submillimeter Array (SMA) by \cite{Krips11}.  We estimate the mass outflow rate as 
$d{M}_{out}/dt=3\times v_{max}\times M_{out}/R_{out} \simeq$108 M$_{\odot}$ yr$^{-1}$ , and  the kinetic luminosity from eq. (\ref{Lkin}) with $v_{out}=v_{max}$. We derive $L_{kin}$=1.5$\times$10$^{42}$ erg/s.

 The  kinetic luminosity of the outflow indicates that AGN activity rather than  star formation  is the likely powering source of the outflow.  Indeed,  the  star formation rate for the circumnuclear region out to a radius  $R\simeq$140 pc is   SFR$ \simeq $1 M$_{\odot}$/yr \citep{Esquej14}. This corresponds  to a supernovae rate  $\nu_{SN}\simeq$0.01 yr$^{-1}$ for a Kroupa initial mass function \citep{Kroupa01}. Assuming a typical kinetic energy from a supernovae explosion of E$_{SN}$=10$^{51}$ erg,  the kinetic luminosity associated to star formation is L$_{kin}$=$\nu_{SN}$E$_{SN}\simeq$3$\times$10$^{41}$ erg/s.\\ 
The jet power derived from the luminosity at 1.4 GHz  L$_{jet}$=1.8$\times$10$^{43}$ erg/s  \citep{Garcia14}, and the bolometric luminosity of the active nucleus estimated from mid-IR and X-ray bands  L$_{AGN}$=(4.2$\times$10$^{44}$-2.1$\times$10$^{45}$) erg s$^{-1}$ \citep{Bock00, Alonso11,Garcia14, Marinucci16}  indicate that the interaction of the CND gas with either the AGN jet and/or the energy released during accretion of matter onto supermassive black hole are able to drive the molecular outflow.

\subsection{$\gamma$-ray spectrum}\label{gamma_spectrum}
In the very high energy part of the electromagnetic spectrum NGC 1068 was observed by  {\it Fermi}.
Here we present our analysis of \textsc{NGC 1068} {\it Fermi}-LAT data.
The LAT instrument \citep{Atwood09}  detects $\gamma$--rays in the energy
range extending from 20 MeV to more than 300 GeV, by measuring their
arrival time, energy and direction. We analyzed \textsc{NGC 1068} {\it Fermi}-LAT
data collected from August 8th, 2008 to February 28th, 2016 in the
energy range 100 MeV--300 GeV. We used the \textsc{Fermi Science
  Tools} (version v10r0p5) to produce and analyse cleaned and filtered
event files, following the data analysis thread provided by the {\it Fermi}
Science Support Center\footnote{http://fermi.gsfc.nasa.gov/ssc/data/analysis/scitools/}. We
considered the Pass 8 data (evclass$=$128 \& evtype$=$3) and only
events with zenith angles $<$ 90$^{\circ}$ to minimize the contamination
from the Earth Limb. We filtered the data using good time intervals
generated using the task \textsc{gtmktime}, with expression
recommended by the LAT team of (DATA QUAL$>$0)\&\&(LAT
CONFIG$==$1). We adopted the ``P8R2 SOURCE V6'' instrumental response
function (IRF).

The events are selected from a 40$^{\circ}$ $\times$40$^{\circ}$
squared region of interest (ROI), centered on \textsc{NGC 1068}. We
used the binned maximum likelihood method to obtain the \textsc{NGC
  1068} spectral energy distribution. Our model includes all known
sources in the {\it Fermi}-LAT 4-Year Point Source catalogue \citep[3FGL,][]{Acero15}, the diffuse Galactic
background (gll\_iem\_v06.fits) and diffuse extragalactic component
(iso\_P8R2\_SOURCE\_V6\_v06.txt). The model file was generated with
make3FGLxml.py6. We left free to vary only the brightest and variable
sources in the ROI; moreover, the make3FGLxml.py script automatically
adds 10$^{\circ}$ to the ROI, to account for sources that lie outside,
but which may provide a photon contribution to our data. 
We also included a new source, not included in the 3FGL, located at a distance
less than 4$^{\circ}$ from \textsc{NGC 1068},  centered on the peak of the LAT  100 MeV--100 GeV emission, modelled with a
power--law model with free normalization and photon spectral
index. This source is a factor of $\sim$3 brighter than \textsc{NGC 1068}  in the 100-300 MeV energy range,  with $\sqrt(TS)\sim11$\footnote{The TS value is
  defined as TS$=-2(L0- L1)$, where L0 and L1 are the logarithmic
  maximum likelihood values of null hypothesis and tested model
  including the target source. The square root of the TS is
  approximately equal to the detection significance for a given
  source.} and could be the counterpart of the 5BZQJ0239$+$0416 QSO RLoud
flat--spectrum radio quasars \citep{Massaro15}, located at a
distance $\sim$15\amin from its LAT position.
We found that \textsc{NGC 1068} is fainter at energies 100--300 MeV
than in the 3FGL catalogue. We verified that this is due to
the addition of the new source. In fact, owing to the large low energy
PSF (i.e. 95\% PSF is $\sim 10^{\circ}$ at $\sim$100 MeV), the 3FGL
flux is probably overestimated, including part of the emission of the
new source. \\
At energies $>$10 GeV we found a spectral shape different
from that of the 3FGL. This discrepancy is ascribed to the splitting of the energy bin, indeed  using the same single bin (10--100
GeV) we reproduce the 3FGL results. Our analysis indicates a significant ($\sim$ 5 $\sigma$) $\gamma$-ray emission   at energies 10--30 GeV, while in the 30--100 GeV energy range we obtain only an upper limit.
The 3$\sigma$ upper limit is computed as follows: if T is the total
number of counts measured at the position of a source, B are the
expected background counts, and X are the unknown counts from the
source, the 3$\sigma$ (99.7\%) upper limit on X can be defined as the
number of counts X that gives a 0.3\% probability to observe T (or
less) counts. Applying the Poisson probability distribution function,
X is therefore obtained by iteratively solving for different X values
in the following equation:

\begin{equation}
0.003=e^{-(X+ B)}\sum_{i=0}^T { (X+ B)^i \over i!}
\end{equation}
(see e.g., \citealt{Narsky00})

We collected the counts T from a region centered on \textsc{NGC
  1068} with 1$^{\circ}$ radius, which contains more than 95\% PSF,
and estimated the B expected background counts in source free
neighboring regions. Finally we evalutated the flux from the counts
applying the vignetting corrected exposure time stored in the exposure
maps.\\
 The spectrum obtained from this anaysis as well as the 3FGL spectrum are shown in figure  \ref{bestfit} (see also table 1).

\begin{center}
\begin{table*}[h!]
 \begin{tabular}{|c|c|c|c|c|c|c|}
 \hline
   F$_{0.1-0.3 GeV}$ & F$_{0.3-1 GeV}$  & F$_{1-3 GeV}$ & F$_{3-10 GeV}$ & F$_{10-30 GeV}$ & F$_{30-100 GeV}$\\   
    (10$^{-12}$ erg cm$^{-2}$ s$^{-1}$) &  (10$^{-12}$ erg cm$^{-2}$ s$^{-1}$)   & (10$^{-12}$ erg cm$^{-2}$ s$^{-1}$)   & (10$^{-12}$ erg cm$^{-2}$ s$^{-1}$)   &  (10$^{-12}$ erg cm$^{-2}$ s$^{-1}$)  &  (10$^{-12}$ erg cm$^{-2}$ s$^{-1}$)  \\ \hline
 1.57$\pm$0.37 & 1.46$\pm$0.19& 0.98$\pm$0.15 & 0.54$\pm$0.16 & 1.12$\pm$0.41 & $<$ 1.90 \\ \hline
\end{tabular}
\caption{\textsc{NGC
  1068} spectral informations derived from the {\it Fermi}-LAT analysis presented in this paper (Section \ref{gamma_spectrum}).}
\end{table*}
\end{center}

\section{Theoretical modelling}\label{model}

\subsection{Accelerated particle spectra}\label{spectra}
\begin{figure}[h!]
\begin{center}
\includegraphics[width=8 cm]{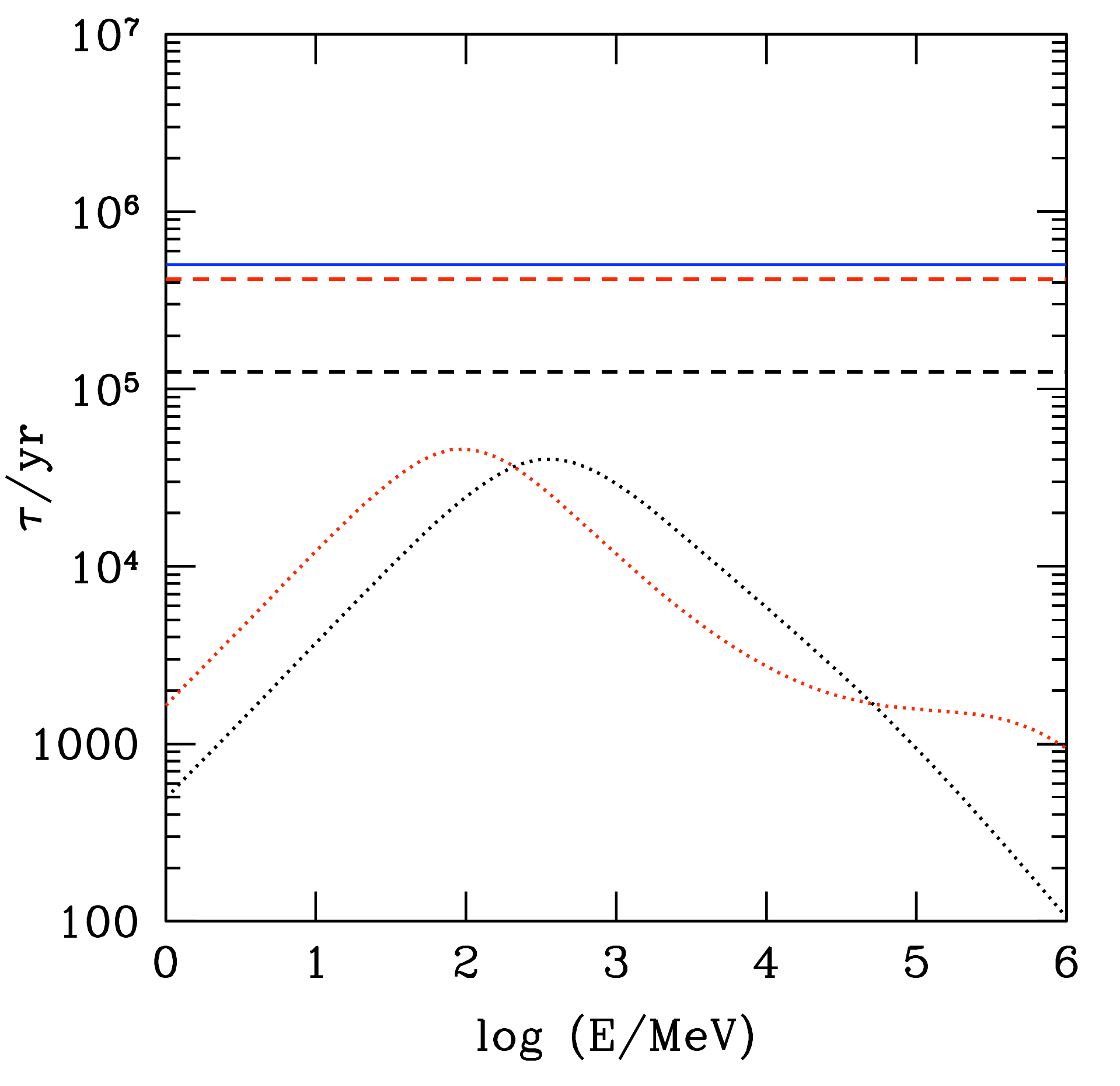}
\caption{Residence times  of CR protons (dashed) and electrons (dotted) in the CND of \textsc{NGC 1068} as a function of particle's energy. 
 Red lines $n_H$=120 cm$^{-3}$, B$_{ISM}$=100$\mu$G, L$_{AGN}$=2.1$\times$10$^{45}$ erg/s, U$_{rad}$=3.7$\times$10$^4$ eV cm$^{-3}$, $R_{shock}$=100 pc, and $v_{shock}$=200 km/s; black lines:  $n_H$=400 cm$^{-3}$, B$_{ISM}$=350$\mu$G, L$_{AGN}$=4.2$\times$10$^{44}$ erg/s, U$_{rad}$=7.3$\times$10$^3$ eV cm$^{-3}$, $R_{shock}$=100 pc, and $v_{shock}$=200 km/s. 
 The solid blue line indicates the  time-scale  of the AGN-driven molecular outflow. }
\label{ep_lifetime}
\end{center} 
\end{figure}

To build-up a model that predicts the $\gamma$-ray and radio spectra, we must first calculate the CR spectra. \\
CR particles are subject to a number of energy loss processes which cause distortions of their injection spectra as they propagate through the ISM.  
In the case of relativistic protons, pions production through inelastic collisions with ambient protons is the dominant loss mechanism.  The collisional energy loss time-scale is given by:
\begin{equation}\label{tau_pp}
\tau_{pp} \approx 5 \times 10^7 yr \left(\frac{n_H}{cm^{-3}}\right)^{-1}
\end{equation}
where $n_H$ is the ISM  gas density. \\
Protons can also escape from the acceleration region. The time-scale on which CR particles are advected in the AGN-driven outflow is:
\begin{equation}\label{tau_flow}
\tau_{out}=\frac{R_{out}}{v_{out}}.
\end{equation}   
From the above equations we compute the residence time of CR protons $\tau_{p}$  as:
\begin{equation}\label{tau}
\tau_{p}=(\tau_{pp}^{-1}+\tau_{out}^{-1})^{-1}.
\end{equation}
A galaxy becomes a proton calorimeter  when the residence time approximately equals the collisional energy loss time-scale ($F_{cal}\equiv \tau_{p}/\tau_{pp} =$1).\\
For relativistic electrons,  the loss mechanisms involve interactions with matter, magnetic field, and radiation.  
For a gas density $n_H$ the energy loss rate due to bremsstrahlung emission is:
\begin{equation}\label{bremm}
-\left(\frac{dE}{dt}\right)_{brem}=3.66\times 10^{-7} \left(\frac{E}{eV}\right) \left(\frac{n_{H}}{cm^{-3}}\right) \quad eV s^{-1}
\end{equation}  
while the  loss rate  due to ionization is:
\begin{equation}\label{ion}
-\left(\frac{dE}{dt}\right)_{ion}=7.64\times 10^{-15} n_H(3\ln(E/m_ec^2)+19.8) \quad eV s^{-1}.
\end{equation}  
The  synchrotron cooling rate  is given by:
\begin{equation}\label{syn}
-\left(\frac{dE}{dt}\right)_{syn}=\frac{4\sigma_T c E^2 U_{B} }{3(m_ec^2)^2}
\end{equation}
where $m_e$ is the electron mass, $\sigma_T$= 6.65$\times$10$^{-25}$cm$^{-2}$ is the Thomson cross section, and U$_B$=B$_{ISM}^{2}/8\pi$ is the  magnetic energy density  . Finally, the energy loss rate of a relativistic electron by IC scattering in the Thomson approximation is:
\begin{equation}\label{IC_formula}
-\left(\frac{dE}{dt}\right)_{IC}=\frac{4\sigma_T c E^2 U_{rad} }{3(m_ec^2)^2}
\end{equation}
where U$_{rad}$=L$_{AGN}/4\pi cR_{out}^2$ is the radiation energy density at the location of the shock (R$_{out}$), and L$_{AGN}$ is the AGN luminosity\footnote{ The IR luminosity derived from the  circunuclear SFR and the SFR-L$_{IR}$ relation \citep{Kennicutt98,Rieke09} is  L$_{IR} \simeq$3.3$\times$10$^{43}$erg s$^{-1}$. Thus within the CND the energy density of the AGN radiation field overwhelms that of the stellar component.}.
 The full Klein-Nishina cross section is used to compute the IC cooling time-scale  and the photon emission (see fig. \ref{ep_lifetime} and Section \ref{IC}).  Klein-Nishina effects imply a significant reduction of the CR electron energy loss rate compared to the Thomson limit (see e.g. \citealt{Schlickeiser10}).\\
The residence time of CR electrons with energy $E$ is therefore:
\begin{equation}
\tau_e=\frac{E}{(dE/dt)} 
\end{equation}
where $(dE/dt)$  is the sum of the processes described in eq. (\ref{bremm})-(\ref{IC_formula}). \\
Figure \ref{ep_lifetime} shows the residence times of CR protons and electrons in the CND of \textsc{NGC 1068} as a function of the particle's energy. The  time-scale  on which CR particles are advected in the  AGN-driven molecular outflow ($\tau_{out}$) is also shown.  The latter is comparable to the proton residence time, but much longer than the electron residence time. This implies that CR protons either interact with the CND gas or escape in the outflow, whereas CR electrons predominantly stay and cool in the CND.\\

The evolution of the CR energy spectrum as the particles diffuse from their source under conditions of energy losses is given by the diffusion-loss equation.  If the time-scale for particle acceleration is smaller than all other time-scales (energy loss or escape) than the system will be in a steady state.
The steady-state diffusion-loss equation for CRs  with no spatial and temporal dependence is  \citep{Longair11}:
\begin{equation}\label{diffusion_eq}
\frac{N(E)}{\tau_{life}(E)} - \frac{d}{d E}[b(E)N(E)]-Q(E)=0
\end{equation}
where E is the total energy, $N(E)$ is the CR spectrum, $Q(E)$ is CR source function , $b(E)=-(dE/dt)$ is the cooling rate of individual CR,  and $\tau_{life}(E)$   is the lifetime to diffusive or advective escape from the system. \\
Since pion production can be identified as a catastrophic loss \citep{Torres04,Lacki13}, the final proton spectrum can be obtained by solving eq. (\ref{diffusion_eq}) with $b(E)\rightarrow $0:
\begin{equation}\label{p_spectrum}
N_p(E) \simeq Q_p(E)\tau_p
\end{equation}
where $\tau_{p}$ is the residence time of CR protons given in eq. (\ref{tau}). Note that the latter neglects diffusion losses. The role of diffusive escape constitutes a source of uncertainty in the computation. On the other hand, energy-dependent diffusion losses soften the $\gamma$-ray spectrum. 
The hard $ \gtrsim $ GeV spectrum found for \textsc{NGC 1068} suggests that the energy losses  due to either hadronic interactions or advection dominate over diffusion losses. In the following we will neglect diffusion losses for both CR protons and electrons.\\
Since  escape (advective and diffusive) is negligible  for CR electrons  ($\tau_{life}(E)\rightarrow$0) the final spectrum will have the form: 
\begin{equation}
N_e(E) \propto \frac{Q_e(E)\tau_e(E)}{p-1}
\end{equation}
if the spectrum of the injected electrons has the form $Q(E)\propto E^{-p}$.\\

It is generally believed that particle acceleration in strong shock waves, often referred as diffusive shock acceleration (DSA),  is the primary mechanism producing energetic particles in astrophysical shocks (but see \citealt{Vazza15,Vazza16,vanWeeren16} for results showing that DSA  has difficulties in explaining the observed emissions of particles accelerated in galaxy cluster shocks).
 DSA could result in the production of a  power-law CR population with a power-law index $p\simeq$2 \citep{Bell78,Bell78b,Blandford78,Drury83}, that extends to energies as high as is permitted by various loss processes.
Here we assume that DSA is effective in AGN-driven shocks, so we assume for the source functions  the form:
\begin{equation}
Q_p(E)=A_pE^{-p}\exp\left[-\left(\frac{E}{E_{max,p}}\right)\right]
\end{equation}
and 
\begin{equation}
Q_e(E)=A_eE^{-p}\exp\left[-\left(\frac{E}{E_{max,e}}\right)\right]
\end{equation}
where the normalization constants $A_p$ and  $A_e$ are determined by the total energy supplied to relativistic protons and electrons at the shock respectively, and $E_{max,p}$ and $E_{max,e}$ are the maximum energies of accelerated protons and electrons.\\
 As AGN-driven outflow is the assumed driver of CR acceleration, the CR particle spectrum must be related to the total energy  input from the AGN:
\begin{equation}\label{Lkin_norm}
\int_{E_{min}}^{E_{max}} N(E) E dE = \eta  \int_{\tau_{out}-\tau}^{\tau_{out}}  L_{kin}(t) dt
\end{equation}
here $L_{kin}$ is the kinetic luminosity of the outflow,  $\eta$ is the fraction of the  kinetic energy transferred to CR particles, $\tau_{out}$ is the outflow time-scale, and $\tau$ is the residence time of CR particles. 
The physics  governing  the acceleration of particles  in non-relativistic AGN-driven outflows is not well known, in particular  we have no indication of the  particle's acceleration efficiency. A general indication 
can be obtained by viewing AGN-driven outflows as a SNR analogue.
The comparison between the $\gamma$-ray and radio emissions from SNR and the  kinetic energy supplied by  supernova explosions suggest 
$\eta_p\simeq$(0.1-0.2) for CR protons and $\eta_e\simeq$(0.01-0.02) for CR electrons \citep[][and references therein]{Keshet03,Thompson06,Tatischeff08,Lacki10}.\\

The maximum energy of accelerated particles depends on  the age or size of the accelerator, on the particle energy-loss processes, and on the scale $\lambda$ of the particle diffusion length. The expressions  for the maximum achievable energy in the SNR context are  given in \cite{Reynolds08}:
 \begin{equation}\label{Emax_age}
E_{max,age}=0.5v_{s,8}^2t_{age,3}B_{\mu G}f^{-1}  TeV
\end{equation} 
\begin{equation}\label{Emax_loss}
E_{max,loss}=100v_{s,8}(fB_{\mu G})^{-0.5}   TeV
\end{equation} 
\begin{equation}\label{Emax_esc}
E_{max,esc}=10B_{\mu G} \lambda_{17}  TeV
\end{equation} 
where $v_{s,8}$ is the shock velocity in units of 10$^8$ cm/s, $t_{age,3}$ is the age of the accelerator in units of 10$^3$ yr, $ B_{\mu G} $ is the magnetic field strength in units of $\mu$G, $\lambda_{17}$ is the diffusion length in units of 10$^{17}$ cm, and $f$ parametrizes diffusion effects ($ f\simeq $1 for Bohm diffusion).

\subsection{The physics of AGN-driven galactic outflows}\label{outflow_physics}
The normalizations of the CR proton and electron spectra (eq. \ref{Lkin_norm}) depend on the kinetic energy injected into the ISM by the AGN-driven outflow during the residence time of CR particles.
In this context, the most important question regarding the  physics of AGN-driven galactic outflows is whether or not the outflow kinetic luminosity is constant in time. \\
Theoretical models postulate that the galactic non-relativistic outflows are produced when wide-angle fast winds ($ v\simeq $0.1-0.3 $c$) shock against the galaxy ISM (e. g. \citealt{King03,King03_2,King11,Lapi05,FGQ12,Zubovas12,Zubovas14}).  These mildly relativistic winds  are observed in the central regions of AGN (e.g. \citealt{Chartas02,Pounds03,Reeves03,Tombesi10,Tombesi15}), and likely originate from the acceleration of gas particles in the accretion disk around super massive black holes by the AGN radiation field. \\
The fast wind impact implies a reverse shock that slows the wind and injects energy into the ISM. The shocked wind acts like a piston sweeping up the ISM. The supersonic swept up gas  drives  a forward shock into the ISM, which is separated from the reverse shock by a contact discontinuity. \\
The cooling of the shocked fast wind  determines whether  the  outflow kinetic energy is conserved \citep{King11,FGQ12}.  If the shocked wind gas cools efficiently, most of the pre-shock kinetic energy is lost (usually to radiation), and only its momentum flux is transferred to the ISM (momentum-driven outflow). On contrast,  if the cooling of the shocked wind gas is inefficient, the  shocked wind gas  retains all the mechanical luminosity termalized in the shock, and expands adiabatically pushing the ISM gas away (energy-driven outflow). In this case the momentum flux of the swept-up material in the ISM  increases with time  owing to the work done by the shocked gas. The expected  momentum boost is predicted to be $(dP_{out}/dt)/(L_{AGN}/c)\simeq$10-100  \citep{Zubovas12,FGQ12}, where $(dP_{out}/dt)=(dM_{out}/dt)v_{out}$ is the momentum flux of the large-scale outflow, and $L_{AGN}/c$ is the momentum flux of the fast wind accelerated by AGN radiation pressure. \\
 By comparing the IC cooling time-scale of the shocked wind  and the outflow time-scale  \cite{King03} (see \citealt{King15} for a review) found  that momentum-driven outflows are confined to the nuclear region of galaxies, while energy-driven outflows affect larger scale (but see \citealt{FGQ12} for different results implying  energy-driven  outflow for a wide range of parameters characteristic of real galaxies).\\
Overall, the observations of kpc-scale AGN-driven molecular outflows in ultra luminous infra-red galaxies (ULIRGs)  indicate large values of the  ratio $(dP_{out}/dt)/(L_{AGN}/c)$ strongly suggesting the energy-driven regime for these outflows (e.g. \citealt{Fischer10,Sturm11,Rupke11,Feruglio10,Feruglio15,Cicone14}). However,  for the more compact molecular outflow observed in \textsc{NGC 1068} we are not in the position  to clearly distinguish its nature. Indeed, assuming the  outflow mass rate and velocity derived in Section \ref{CND} we find $(dP_{out}/dt)$=1.4 $\times$10$^{35}$ g cm s$^{-2}$.  The latter is a moderate factor  $\sim$(2-10)  larger than $(L_{AGN}/c)$=(1.4 $\times$10$^{34}$-0.7 $\times$10$^{35}$) g cm s$^{-2}$. 
Thus in the next sections, we will derive the $\gamma$-ray and radio emission by assuming both  an energy-driven  and a momentum-driven outflow. In the former case,  $L_{kin}$  is assumed constant in eq. (\ref{Lkin_norm}),  and  equal to the kinetic luminosity derived from  sub-mm interferometry (see Section \ref{CND}). In the latter  case, $L_{kin}$ varies in time. Since the kinetic luminosity $L_{kin}\propto (dM_{out}/dt)v_{out}^2\propto \rho(R)R^2v_{out}^3$,  in order to estimate  the  total kinetic energy injected into the ISM during the residence time of CR particles from the measured kinetic power we need to know the gas density profile and  the  dynamics of the expansion.\\
To this purpose we adopt the models of \cite{King10} and \cite{Lapi05}  (see also \citealt{Menci08}).\\
 Assuming  an isothermal gas density profile and that all the fast wind energy not associated with the ram pressure is rapidly lost to radiation, \cite{King10} derived the analytic solution of the equation of motion of the momentum-driven shock pattern  using the shell approximation. For the region outside the black hole sphere of influence they found  $R(t)\propto at+bt^{0.5}$+c where the constants $a$, $b$, and $c$ are determined by the black hole mass, the parameters of the galaxy potential, and by the position and the speed of the shell at time $t$=0 (see \citealt{King10}). The above relation  implies  $L_{kin}(t)\propto t^{-1.5}$.\\
Assuming  a power law ambient density profile $\rho(R) \propto R^{-\alpha}$ ($ 2\le\alpha\le2.5$),  and the Rankine-Hugoniot boundary conditions in terms of the Mach number, \cite{Lapi05} derived self-similar  solutions  of the partial differential equations describing the gas flow. They found  $R(t)\propto t^{2/\alpha}$ implying an outflow power  $L_{kin}\propto t^{(10-5\alpha)/\alpha}$.\\
These  models  correspond to kinetic energies injected into the ISM during the outflow time-scale   larger up to a factor of $ \sim$2 than those injected in the energy-conserving case.

\subsection{Non-thermal emission produced by accelerated particles}\label{emission}

\subsubsection{$\gamma$-ray emission from neutral pion decays}\label{gamma}
CR protons can inelastically scatter off protons in the ISM to produce neutral and charged pions. Neutral pions decay into two $\gamma$-rays: $\pi^{0} \rightarrow \gamma + \gamma$; while charged pions decay into secondary electrons and positrons and neutrinos: $\pi^{+} \rightarrow \mu^{+} + \nu_{\mu}$ and $\mu^{+} \rightarrow e^{+}+\nu_{e} +\overline{\nu}_{\mu}$; $\pi^{-} \rightarrow \mu^{-} + \overline{\nu}_{\mu}$ and $\mu^{-} \rightarrow e^{-}+\overline{\nu}_{e} +\nu_{\mu}$.\\
To calculate the hadronic component  of the  $\gamma$-ray spectrum and the energy spectra of secondary particles (electrons, positrons and neutrinos) we use the parametrizations derived by  \cite{Kelner06}. These analytical approximations, which  are based on simulations of proton-proton interactions, provide very good accuracy over the energy range of parent protons above 100 GeV.
At lower energies the spectra   are calculated using the $\delta$-functional approximation \citep{Aharonian00}.

\subsubsection{$\gamma$-ray emission  from inverse Compton}\label{IC}
The leptonic component of the  $\gamma$-ray spectrum consists of IC scattering of the AGN radiation and non-thermal bremsstrahlung.
The spectrum of photons generated per unit time due to IC scattering  is \citep[see e.g.][]{Blumenthal70}:
\begin{equation}\label{icm2}
d\dot{N}_{\gamma}/dE_{\gamma}=\int d\epsilon n_{ph}(\epsilon) \int dE_{e} W(E_{e},\epsilon,E_{\gamma})N_{e}(E_{e})
\end{equation}
where
\begin{equation}\label{icm3}
n_{ph}(\epsilon)=\frac{1}{4\pi c R_{out}^2 \epsilon}\frac{dL_{AGN}}{d\epsilon}
\end{equation}
is the differential input photon number density at the location of the shock. To calculate the output spectrum we use the synthetic spectrum of a type 1 AGN as computed by \cite{Sazonov04} normalized to the bolometric luminosity of NGC1068. In eq. (\ref{icm2}) $N_e$ represents the combined primary and secondary electron/positron spectrum from pion decay, and
\begin{equation}\label{icm4}
W(E_e,\epsilon,E_\gamma)=
\frac{8\pi r_e^2 c}{E_e\,\eta}\left[ 2q\,\ln q+(1-q)\left(1+2q+
\frac{\eta^2q^2}{2\,(1+\eta q)}\right)\right],
\end{equation}
is the scattering cross-section, taking into account Klein-Nishina effects, with  $\eta=\frac{4\,\epsilon E_e}{m_e^2 c^4}$ and $q=\frac{E_\gamma}{\eta\,(E_e-E_\gamma)}$.

\subsubsection{$\gamma$-ray emission  from bremsstrahlung}

In the case of production by bremsstrahlung the spectrum of photons generated per unit time  is \citep{Stecker71}:
\begin{equation}\label{brem1}
d\dot{N}_{\gamma}/dE_{\gamma}=cn_H\sigma_{brem}E_{\gamma}^{-1}\int dE_{e} N_{e}(E_{e})
\end{equation}
where $\sigma_{brem}=3.38\times10^{-26}$ cm$^2$, and $N_e$ represents the combined primary and secondary electron/positron spectrum.

\subsubsection{Radio emission from synchrotron}
The majority of the radio flux comes from non-thermal synchrotron emission. 
We calculate the synchrotron emission following the standard formula from \cite{Rybicki79}.
We calculate the radio spectra due to  synchrotron emission by primary electrons and secondary electrons/positrons from pion decay. 
The radiation spectrum of a power law electron energy distribution $N_e(E)\propto E^{-p}$ may be expressed in terms of the magnetic field $B_{ISM}$ and the frequency $\nu$ as $F_\nu\propto B_{ISM}^{(p+1)/2} \nu^{-(p-1)/2} $. Thus the spectral shape is determined by the shape of the electron distribution: if $-p$ is the spectral index of the electron energy spectrum, the spectral index of the synchrotron emission   is $-(p-1)/2$.

\section{Results}\label{results}

Here we derive the non-thermal emission produced by CR particles accelerated in the AGN-driven outflow observed in  \textsc{NGC 1068} and we compare it with  the  $\gamma$-ray and radio spectra.\\

 We compare the predicted $\gamma$-ray emission with  the $\gamma$-ray spectrum from the  3FGL  catalogue \citep{Acero15}, and with the $\gamma$-ray spectrum  obtained from our analysis of the 
 {\it Fermi}-LAT data (Sect. \ref{gamma_spectrum}). The  {\it Fermi}-LAT  observations do not spatially resolve the  $\gamma$-ray emitting region. Therefore, we suppose that all the $\gamma$-ray flux is related to the  inner $ \sim $ 100 pc of the galaxy. Conversely, the radio observations are spatially resolved. However within 100 pc the presence of the  radio jet hampers the identification of any emission not originating from it or the compact nucleus.  The current radio observations provides therefore  only limited constraints on the cosmic ray population outside the AGN jets \citep[see also][]{Yoast14, Eichmann15}.  
In the following analysis  we consider the magnetic field derived from the radio observations as only an upper limit to the magnetic field strength in the shock region.\\

 We derive the $\gamma$-ray and radio emissions for AGN-driven outflows in both the  energy-driven and  momentum-driven  scenarios (Section \ref{outflow_physics}). We assume for the outflow radius,  velocity, and kinetic power the values  obtained from sub-mm interferometry:  $R_{out}$=100 pc, $v_{out}$=(100-200) km/s, $L_{kin}$=(0.5-1.5)$\times$10$^{42}$ (see Section \ref{CND}); and for the proton and electron acceleration efficiencies the SNR values: $\eta_p$=(0.1-0.2), and $\eta_e$=(0.01-0.02) \citep{Lacki10}. We vary parameters such as the spectral index ($p$), the AGN bolometric luminosity  ($L_{AGN}$), the density experienced by CR particles ($n_{H}$), and  the magnetic field (B$_{ISM}$). The spectral index  is restricted to the value $p<$2.2  since larger values are ruled out by the combination of {\it Fermi}-LAT and IceCube observations  \citep{Murase13}. 
We vary the AGN bolometric luminosity within the  range   L$_{AGN}$=(4.2$\times$10$^{44}$-2.1$\times$10$^{45}$) erg s$^{-1}$, and  the density experienced by CR particles 
in the range $n_{H}$=(115-460) cm$^{-3}$ (see Sect. \ref{CND}). In this way we are assuming that the density experienced by CR particles is  equal to  the average density of the CND. However,  CR particles do not necessarily propagate through gas with the average ISM  density . This may be because the ISM is clumpy and the accelerated particles  favour  high-density clumps or they  propagate along path of least resistance. Thus, we also consider a simple calorimetric hadronic model  in which CR protons loose almost  all of their energy to hadronic collision before escaping  ($F_{cal}$=1). This condition could be satisfied if $\tau_{pp} \ll \tau_{out}$  and hence if $n_{H}\gtrsim$10$^{4}$ cm$^{-3}$.  Finally,  for the magnetic field  we assume values bracketed by a minimum value that is given by the volume average magnetic field strength of the CND (see Section \ref{CND}) and a maximum value that is inferred from the radio observations.\\
 The results are shown in figure \ref{bestfit}. We find that, in the standard SNR acceleration theory, even selecting parameters to maximize the $\gamma$-ray emission, the predicted spectrum  is lower  than the observed data  by a factor of $ \sim $2  at energies $ E\gtrsim $1 GeV, and by a factor  of $ \sim $10 at $ E\simeq $0.1 GeV.\\
In order to match the observed $\gamma$-ray spectrum acceleration efficiencies $\eta_p\geq$0.3 and $\eta_e\geq$0.1 are required.  A strong coupling between the molecular medium in the CND and CR protons produced in the outflow shocks ($F_{cal}\geq$0.5) is also necessary (see fig. \ref{bestfit}).\\
In the following we study the effect of changing each of the free quantities within the above ranges.

\subsection{Energy-driven and momentum-driven  outflow}

The degree of which the observed outflow is energy-driven or momentum-driven  influences the normalization of the CR particle spectrum (eq. \ref{Lkin_norm}).
The difference in normalization in the two outflow regimes is maximized when the residence time of CR particles is equal to the outflow time-scale.  Since the residence time of CR electrons in the CND is much smaller than  the outflow time-scale  (see fig. \ref{ep_lifetime}), the leptonic components of the $\gamma$-ray spectrum and the radio spectrum are not strongly  influenced by the nature on the outflow.\\ 
The residence time of CR protons  is nearly equal to the  outflow time-scale for ISM density $n_H=$120 cm$^{-3}$ and  decreases  with increasing  ISM density. Thus only for low $n_H$ values we find a significant difference in the two  outflow regimes.

\subsection{CR parameters:  $p$, $\eta_p$, $\eta_e$}

The slope of the hadronic component of the $\gamma$-ray spectrum reflects the slope of the underlying CR proton population $p$.  A steeper primary CR proton spectrum correspond to a steeper $\gamma$-ray spectrum at energies $ \gtrsim $ 1 GeV.\\
The efficiencies  $\eta_p$ and  $\eta_e$ are the normalizations of the injected primary CR protons and electrons respectively (eq. \ref{Lkin_norm}). Larger (lower)   $\eta_p$ and  $\eta_e$  linearly increase (decrease) the hadronic and leptonic spectral components without changing the spectral shape.

\subsection{ISM density}\label{ISM}
The ISM effective density experienced by CR particles determines the efficiency of hadronic losses $F_{cal}$. An increase in  $F_{cal}$ augments the  $\gamma$-ray emission and the number of secondaries (electrons, positrons and neutrinos) from pion decays.  The ISM density also regulates  the importance of bremsstrahlung and ionization losses for CR electrons. \\
Low  density models can be distinguished observationally from the high density models by $\gamma$-ray emission below $ \sim $100 MeV: models with low $n_H$ are dominated by IC emission, whereas models with high $n_H$ by bremsstrahlung emission.\\

\subsection{Magnetic field}
\begin{figure*}[h!]
\begin{center}
\includegraphics[width=9.4 cm]{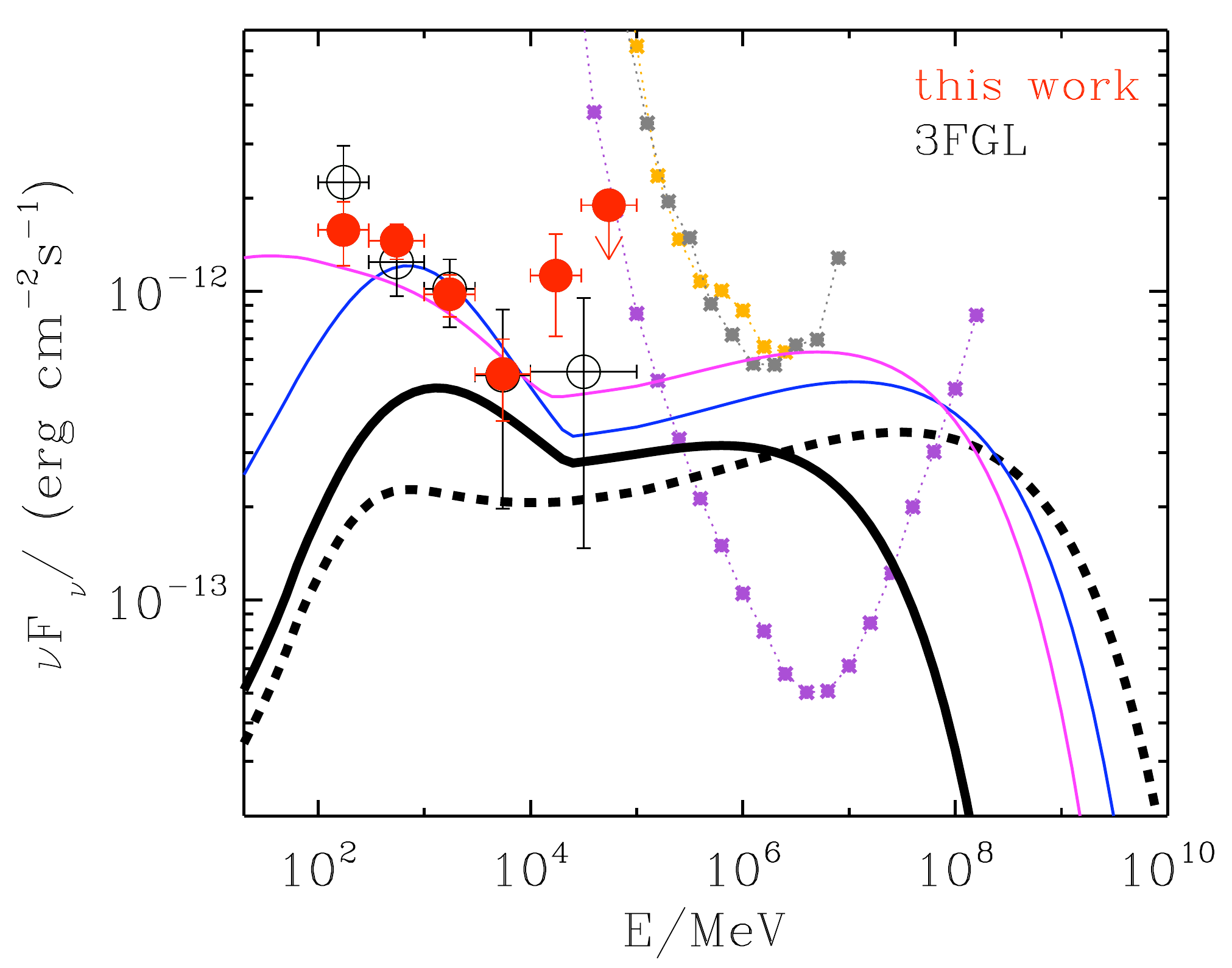}
\includegraphics[width=8.9 cm]{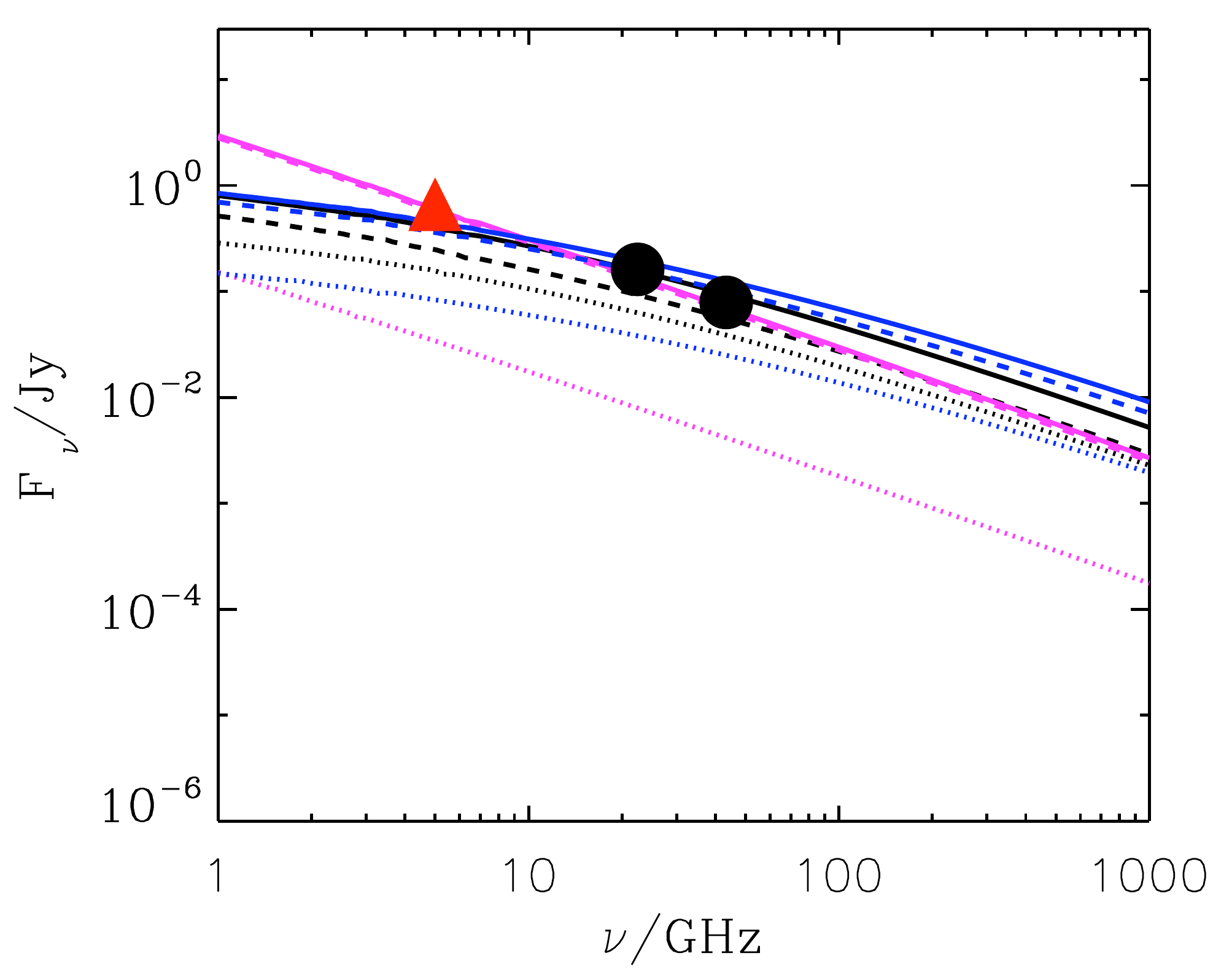}
\caption{$\gamma$-ray and radio spectra for \textsc{NGC 1068}. AGN-driven outflow parameters are set at  $R_{out}$=100 pc, $v_{out}$=200 km/s, and $L_{kin}$=1.5$\times$10$^{42}$ erg/s;  black lines: L$_{AGN}$=4.2$\times$10$^{44}$ erg/s, $n_H$=10$^4$ cm$^{-3}$, $F_{cal}=1$, $p$=2,  $\eta_{p}=0.2$, $\eta_{e}=0.02$, B$_{ISM}$=30 $\mu$G (solid) and B$_{ISM}$=2 mG (dashed); magenta line: L$_{AGN}$=2.1$\times$10$^{45}$ erg/s, $n_H$=120 cm$^{-3}$, $F_{cal}=0.5$, $p$=2,  $\eta_{p}=0.5$, $\eta_{e}=0.4$,  B$_{ISM}$=250 $\mu$G; blue line: L$_{AGN}$=4.2$\times$10$^{44}$ erg/s, $n_H$=10$^4$ cm$^{-3}$, $F_{cal}=1$, $p$=2, $\eta_{p}=0.3$, $\eta_{e}=0.1$,  B$_{ISM}$=600 $\mu$G. $\gamma$-ray spectrum: violet dotted line: expected CTA-south differential sensitivity, from simulated data (5$\sigma$ per energy bin -- 5 per decade -- in 50 hours, zenith angle=20 deg\protect\footnotemark), grey dotted line: measured MAGIC differential sensitivity (5$\sigma$ in 50 hours at low zenith angles $<$30 deg, \citealt{MAGIC_Sens}), orange dotted line: measured H.E.S.S. II differential sensitivity (5$\sigma$ in 50 hours, zenith angle=18 deg, \citealt{Holler15}).  Data points: {\it Fermi}-LAT spectrum from our analysis (red filled circles), and from the 3FGL catalogue (\citealt{Acero15}, black open circles).  Radio spectrum: primary electrons emission (dashed line), secondary electrons/positrons emission (dotted line),  total emission (solid line). Data points: total radio emission from $\simeq$200 pc region (circles, \citealt{Sajina11}),  extra-nuclear radio emission from $\simeq$1.1 kpc region (triangle, \citealt{Gallimore06}).
$^6$ https://portal.cta-observatory.org/Pages/CTA-Performance.aspx
 }
\label{bestfit}
\end{center} 
\end{figure*}

The magnetic field  affects the spectra in several ways. It determines the critical synchrotron frequency $\nu_c\simeq (3E^2eB)/(16m_e^3c^5)$, at a fixed observed frequency a stronger magnetic field implies that we see lower energy electrons. \\
The magnetic field strength  determines the importance of synchrotron cooling relative to IC, Bremsstrahlung, and  ionization losses. If synchrotron cooling dominates over other cooling losses the power law index of the synchrotron radio spectrum changes from -($p$-1)/2 to  -$p$/2, and the radio luminosity approaches a maximum  which is not affected by further increase in  $B_{ISM}$.   The ratio of synchrotron  to IC  losses is equal to $U_B/U_{rad} \propto (B_{ISM}^2 R_{out}^2)/L_{AGN}$ (eq. \ref{syn} and \ref{IC}). At R$_{out}$=100 pc and for L$_{AGN}$=(4.2$\times$10$^{44}$-2.1$\times$10$^{45}$) erg s$^{-1}$ , synchrotron losses start to dominate over IC losses ($ U_B>U_{rad}$) if B$_{ISM} \gtrsim$1.3 mG. \\
Finally, the magnetic field  determines  the maximum energy of CR protons and electrons (eq. \ref{Emax_age}-\ref{Emax_esc}) that in turn determines the high energy cut-off seen in the  $\gamma$-ray  spectrum.
Sensitive TeV observations are  fundamental  to assess the presence of a high-energy cut-off  (see fig \ref{bestfit}). Measuring a high energy cut-off can  constrain in detail acceleration mechanisms operating in the shocks since the maximum energy of accelerated particles depends by  particle escape, age or size of the accelerator,  and  particle energy-loss processes. \\
Sensitive TeV observations are also fundamental to estimate the total non-thermal high energy luminosity of the source, and to constrain the power law index  $p$ of the accelerated CR proton population. To assess the performances of the present and next generation Cherenkov telescopes we show in figure \ref{bestfit}  the differential energy flux sensitivities of the MAGIC telescopes, of the High Energy Stereoscopic System (H.E.S.S.), and of the Cherenkov Telescope Array (CTA)  corresponding to 5$\sigma$ level of confidence per energy bin (5 bins per decade) after 50 hours of effective observation time.

\section{Discussion}\label{discussion}

\subsection{Origin of the $\gamma$-ray emission}

The results presented in the previous section indicate that  the molecular outflow observed in the CND of  \textsc{NGC 1068} 
  provide a contribution to the $\gamma$-ray emission at least comparable to that provided by SNR-driven shocks due to the starburst activity \citep[see][]{Yoast14, Eichmann15}, when the same standard particle acceleration efficiencies are assumed ($\eta_p\approx 0.1-0.2$ for protons, and $\eta_p\approx 0.01-0.02$ for electrons). 
We have shown that  the AGN outflow model can account for the observed $\gamma$-ray emission only when larger acceleration efficiencies are assumed for protons ($\eta_p\geq 0.3$) and for 
electrons ($\eta_e\geq 0.1$), and for calorimetric fractions $F_{cal}\geq 0.5$. \\
At present, there are no clear observational support for such large acceleration efficiencies. In fact, we expect similarity between strong shocks with comparable velocity \citep[e.g.][]{Keshet03}. However, other parameters that determine the physics of the shock waves, like the pre-shock density and the magnetic field,  in the region near the AGN could not be the same than those in the central regions of non active galaxies. \\ 
As for the  calorimetric fraction, we note that  if the molecular outflow is launched when the outflow of ionized gas and/or the  inner jet sweeps the CND disk, they are not in the plane of the CND.
However, a hint in favour of  the strong coupling between CND molecular gas and the CR protons accelerated in the AGN-driven shocks  comes from the chemical analysis of the molecular gas in the CND.  The comparison between the  abundances derived from molecular line surveys of the CND  with those predicted by chemical models indicates that each sub-region of the CND 
could be characterized by three gas phases, two of which are  pervaded by a  CR ionization rate (and/or X-ray activity) enhanced by at least a factor of 10 compared to the Galactic value  (\citealt{Aladro13,Viti14}, see also \citealt{Spinoglio12}).  
In contrast, if shock acceleration is limited  to a narrow zone of the CND, as suggested by preliminary ALMA observations  at $ \sim $0.5$^{\arcsec}$ resolution \citep{Kohno16}, even larger CR proton acceleration efficiencies   are required in order to produce the observed $\gamma$-ray emission.\\
Modelling the chemistry in molecular disks around AGNs is a difficult task since it is affected by the different physical components of the systems, the disk density structure,  and dynamical processes \citep[e.g.][]{Harada13}. Larger number of molecular transitions with high resolution are therefore necessary to properly determine the chemical and physical properties of the CND. \\
 In the case where the $\gamma$-rays  are from  hadronic processes,  the observed $\gamma$-ray emission  can provide  constrains on the CR ionization rate of molecular gas in the CND. However, the CR proton parameters derived by modelling  the $\gamma$-ray emission characterize the proton spectrum at energies $E_p\gtrsim$ 1 GeV  (threshold for pion formation), while the molecular hydrogen is mostly effective ionized by lower energy protons. Although the latter are  also likely accelerated in the shock, in order to derive the expected ionizing CR proton flux one should rely on complex extrapolations of the CR proton spectrum to energies  $E_p <$ 1 GeV  \citep[see e.g.][]{Padovani09,Scuppan12}. \\

An enhanced molecular ionization  of  the CND gas could therefore be an indication that the $\gamma$-ray emission has an hadronic component.
The balance between the hadronic and leptonic contributions to the emission could constrain the origin of  the $\gamma$-ray emission.\\ 
A   leptonic scenario in which the $\gamma$-ray emission of \textsc{NGC 1068} is produced through IC scattering of IR photons from the  relativistic electrons in the misaligned jet at a few tenth of parsecs from the central source was proposed by \cite{Lenain10}. 
 The  observations of \textsc{NGC 1068}  with current and upcoming Cherenkov telescopes are promising to test CR acceleration models in active galaxies.  Indeed,  
the presence of  $\gamma$-ray emission in the very high energy (VHE) band  could provide a hint to assess the presence of hadronic emission.  The leptonic $\gamma$-ray emission hardly extends to $ \gtrsim $ TeV energies owing to the values of the  shock velocity and of the magnetic field strength in the shock region  (see  eq. \ref{Emax_loss},  and  \citealt{Lenain10}). \\
The energy range below  $E \simeq$ 100 MeV is also crucial to  constrain CR acceleration models.
As shown in Section \ref{ISM}, the IC-dominated $\gamma$-ray spectrum  can be distinguished  from that dominated by  bremsstrahlung emission  at  $E \lesssim$ 100 MeV (see figure \ref{bestfit}, and figure 4 of  \citealt{Lenain10}).
 At present this spectral region is not observable, the new $\gamma$-ray missions that are being planned like ASTROMEV\footnote{the sensitivity to be achieved by ASTROMEV for five years of survey   is  $ \simeq $10$^{-12}$-10$^{-11}$ erg cm$^{-2}$ s$^{-1}$ at $E \simeq$ 1-100 MeV.\\ $http://astromev.in2p3.fr$} and ASTROGAM\footnote{the ASTROGAM expected sensitivity  for an effective exposure of 1 year   is  $ \simeq $10$^{-12}$    -10$^{-11}$ erg cm$^{-2}$ s$^{-1}$ at $E \simeq$ 1-100 MeV.\\ $http://astrogam.iaps.inaf.it/scientific\_instrument.html$} may bring a breakthrough in our understanding of $\gamma$-ray  spectra in active galaxies.

\subsection{Neutrino fluxes}

Finally, the "smoking-gun" to demonstrate the presence of a hadronic component in the $\gamma$-ray spectrum is provided by  the detection of neutrino signal from \textsc{NGC 1068}. Indeed, in the  proton-proton interactions  $\sim$2/3 of the pions produced are  charged pions that decay into   muons and neutrinos followed by electrons and positrons and more neutrinos (see Section \ref{gamma}). \\ 
Neutrinos can also be created  in interactions of CR protons with the ambient radiation field.  Neutrinos created in proton-proton and in proton-photon interactions take $ \sim $5\% of the initial CR proton energy.
This means that 1 PeV neutrino is generated by the interaction of $ \sim $20 PeV CR protons. We find that the cooling time $\tau_{p\gamma}$ of 20 PeV CR protons due to the AGN radiation field in the CND of \textsc{NGC 1068}  is a factor (25-100) larger than the proton-proton collisional time-scale. However since $\tau_{p\gamma}$ decreases with increasing proton energy ($\tau_{p\gamma} \propto$ $E_p^{-1}$ if the photon field is a simple power law $n(\epsilon) \propto \epsilon^{-2}$)  the production of $ \gtrsim $ PeV  neutrinos through photo-hadronic processes could not be negligible.\\

\begin{figure*}[h!]
\begin{center}
\includegraphics[width=8.5 cm]{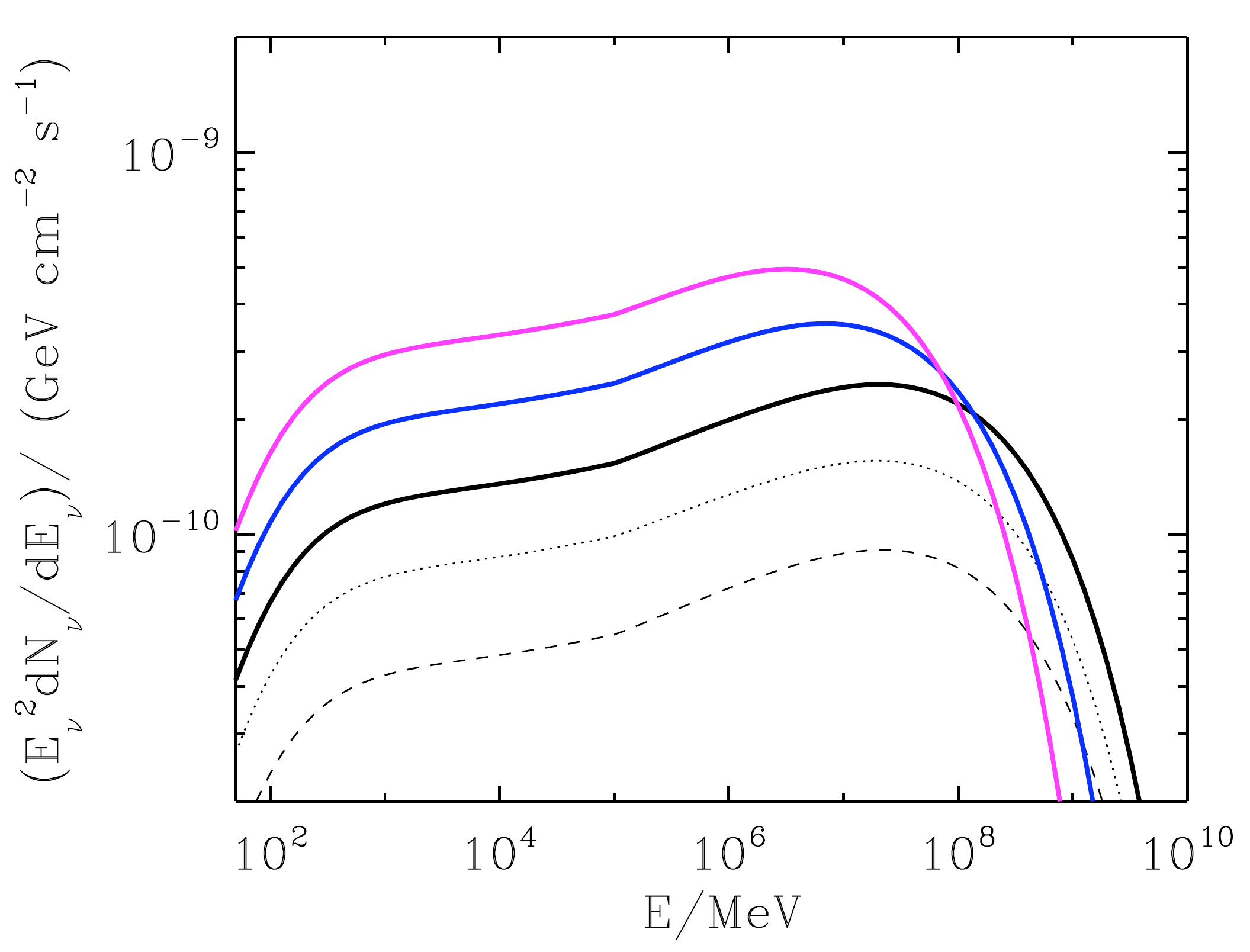}
\includegraphics[width=9 cm]{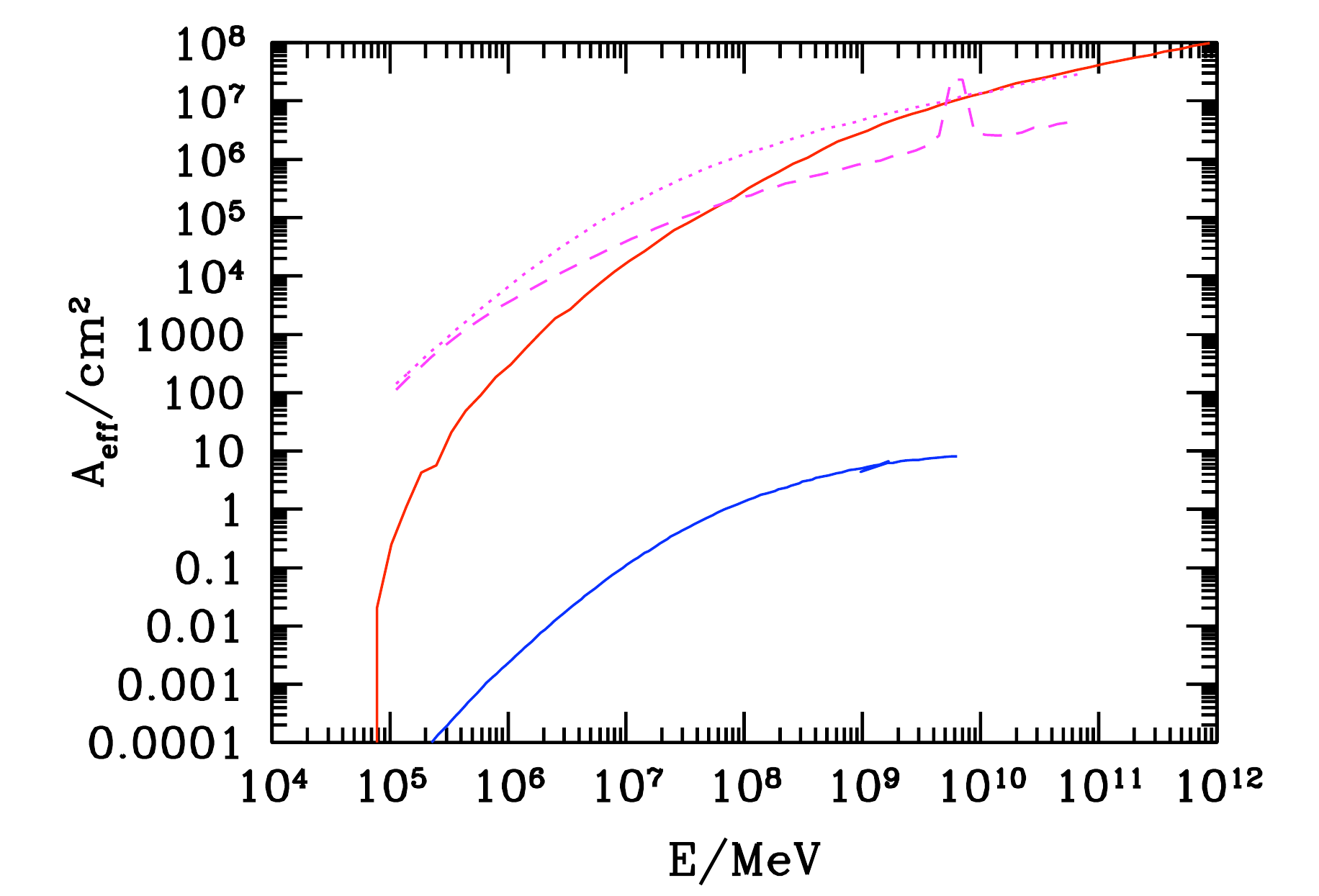}
\caption{Left: neutrino spectra for \textsc{NGC 1068}. Muon neutrino flux (dotted line), electron neutrino flux (dashed line), total  neutrino flux (solid  line). The neutrino fluxes expected from the models in figure \ref{bestfit} are shown with black, blue, and magenta lines. Right:   IceCube effective area (red line),   ANATARES effective area (blue line),  muon neutrino KM3NeT effective area (magenta dotted line), and electron neutrino KM3NeT effective area (magenta dashed line).}
\label{neutrino_flux}
\end{center} 
\end{figure*}

\begin{figure*}[h!]
\begin{center}
\includegraphics[width=18 cm]{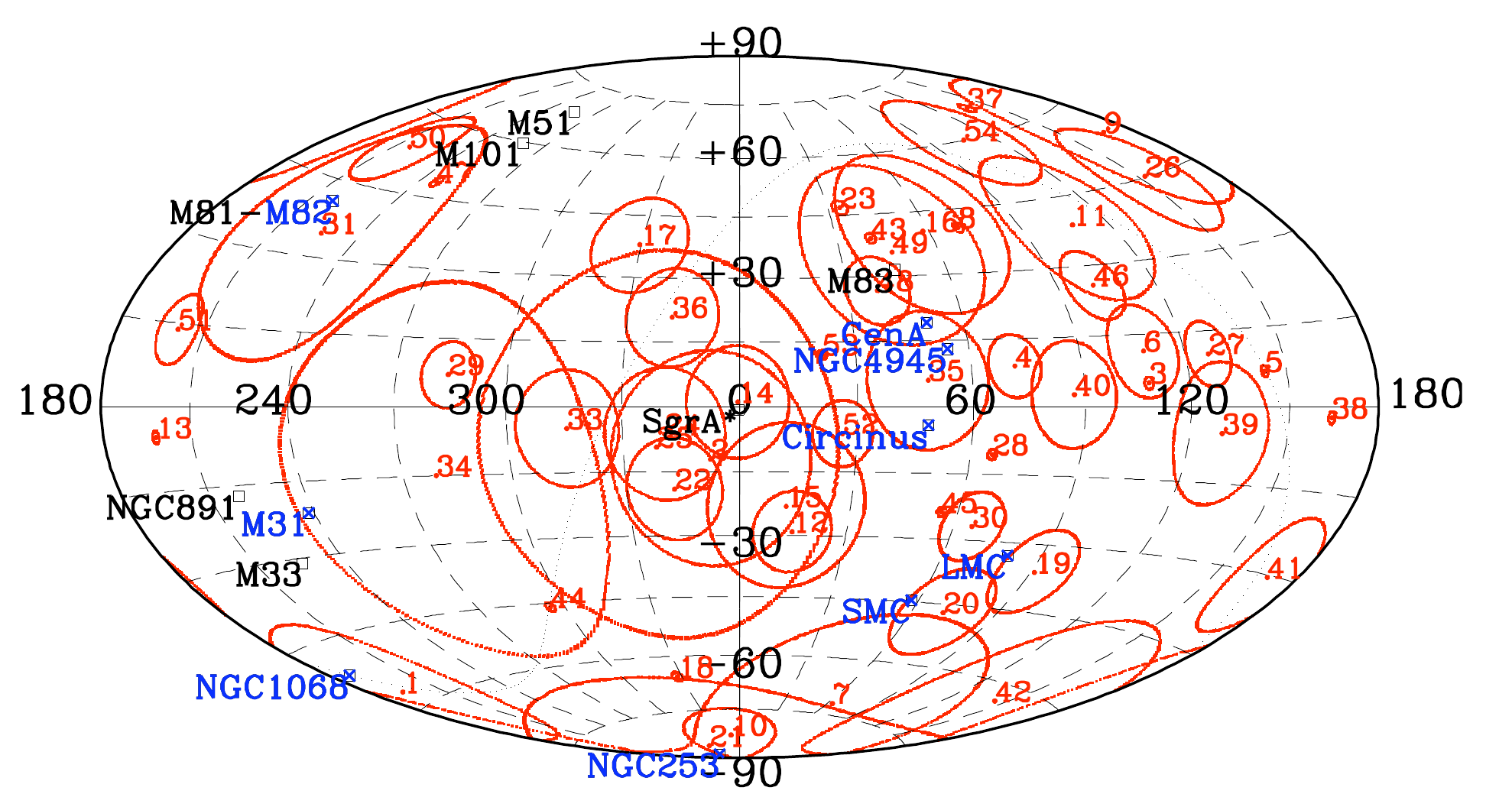}
\caption{Aitoff projection of the IceCube neutrinos in Galactic coordinate system. Red numbered points represent the location of each neutrino event and the surrounding circular areas show the respective median angular error (which includes systematic uncertainties). Black squares are local galaxies with  $v \leqslant$1200 km s$^{-1}$ and IRAS 100 $\mu$m flux  $\geqslant$50 Jy. Blue crossed squares indicate galaxies in the 3FGL catalogue  \citep{Acero15}. \textsc{NGC 1068} is located at the lower left part of the map in correspondence of the neutrino ID 1.}
\label{icecube_events}
\end{center} 
\end{figure*}

Figure \ref{neutrino_flux} shows the neutrino fluxes expected from the models presented in Section \ref{results}.  To calculate the neutrino energy spectra  we follow the procedure described in Section \ref{gamma}. Neutrinos created in proton-proton interactions exhibit energy spectra  that follow the initial CR proton spectrum. \\
In table 2 we report the total number of neutrino events expected in one year of integration time of current and upcoming neutrino detectors. This is obtained from:
\begin{equation}
N_{\nu}=T \times \int \frac{dN_{\nu}}{dE_{\nu}} A_{eff} dE_{\nu}
\end{equation}
where $T$ is the observation time, and $A_{eff}$ is the effective area of \textsc{Antares},  IceCube, and KM3NeT as shown in fig. \ref{neutrino_flux}. The expected number of neutrino events collected  in current neutrino telescopes during 1 year  is $ \lesssim $0.1.
However, our calculation suggests that in the next future, the IceCube' s counterpart on the Northern hemisphere KM3NeT  will be able  to reveal more effectively the neutrino signal from this kind of sources.
 Moreover, thanks to the angular resolution  of the ARCA instrument  of $ \sim $0.2$^{\circ}$ for neutrino events with energy $E  \gtrsim $10 TeV (track-like events, \citealt{KM3Net_perf16}), 
KM3NeT will allow to constrain  effectively the  position of the possible counterparts of neutrino events, thus providing a possible direct test of the AGN outflow model.\\
At present, the analysis of positional coincidence of  neutrino events with known astrophysical objects is a difficult task owing to the  poor angular resolution of current neutrino detectors. 
 Because of these large positional uncertainties  there are no yet  confirmed identifications for  astrophysical sources of high energy neutrino events \citep[e.g.][]{Aartsen14, Adrian16,IceColl16}.\\
Figure \ref{icecube_events} shows the sky map of IceCube neutrino events as well as the  location of  local galaxies (i.e. with recession velocities  $v \leqslant$1200 km s$^{-1}$) with IRAS 100 $\mu$m flux  $F_{IRAS}\geqslant$50 Jy. We also highlight the galaxies detected by {\it Fermi}-LAT. These are non-blazar galaxies with AGN and/or starburst activity in their nuclear regions. The location of \textsc{NGC 1068} is in correspondence  of one of the 54 neutrino events  detected by IceCube in four years of data \citep{Aartsen14}. The fact  that all the  {\it Fermi}-LAT galaxies lie within the acceptance region of a neutrino event suggests an intriguing scenario in which one  may speculate that the $\gamma$-ray emission from particle acceleration in SNR-driven and/or AGN-driven shocks could have hadronic origin.\\
Positional  associations of the  neutrino events detected by IceCube to Seyfert and starburst galaxies in the {\it Fermi}-LAT and IRAS catalogs have been previously reported in the literature \citep{Emig15, Moharana16}. A statistically significant correlation was found between  IceCube  neutrino events and  local starburst galaxies, however, the expected neutrino fluxes derived from the  $\gamma$-rays detected from individual sources  disfavour the  scenario proposed in this paper. We expect that the situation will become clearer when KM3NeT will be operative.\\

\begin{center}
\begin{table}[h!]
 \begin{tabular}{|c|c|c|c|}
 \hline
 & $N_{\nu}$ & $N_{\nu_{\mu}}$ & $N_{\nu_{e}}$ \\ \hline
 \textsc{Antares} &7$\times 10^{-7}$&-&-\\ \hline
 IceCube & 0.1 & - & -\\ \hline
 KM3NeT &-&0.6&0.1  \\
  \hline
\end{tabular}
\caption{ Predicted number of neutrino events in one year of integration time.}
\end{table}
\end{center}

\section{Conclusions}\label{conclusions}
 We compute  the non-thermal emissions produced by CR particles accelerated in the shocks produced by the galactic AGN-driven outflow observed in  \textsc{NGC 1068}.  We find that, within the standard   acceleration theory,  the predicted $\gamma$-ray spectrum  is lower  than the observed data  by a factor of $ \sim $2  at energies $ E\gtrsim $1 GeV, and by a factor  of $ \sim $10 at $ E\simeq $0.1 GeV. This contribution to the $\gamma$-ray emission  is comparable to that provided by the  starburst activity \citep{Yoast14, Eichmann15}. \\
 The analysis presented in this paper indicates that the $\gamma$-ray emission  from \textsc{NGC 1068}  is either entirely produced by leptonic processes - as proposed in the AGN jet model by \cite{Lenain10} - or by processes with acceleration efficiencies of protons and electrons larger than those  commonly assumed in SNR-driven shocks.
The latter interpretation requires either a substantial revision of the standard acceleration theory, or the condition that  AGN-driven shocks are substantially different from  those powered by SNR.\\
The AGN outflow model proposed in this paper can be directly tested by present and forthcoming instruments. 
The observation  of \textsc{NGC 1068}  at TeV energies with present and next generation  Cherenkov telescopes could provide stringent constraints on  CR acceleration models in active galaxies
by the detection of a high-energy cut-off in the $\gamma$-ray spectrum.  Moreover, in the next future the CTA spatial resolution of   $ \sim $ 3 arcmin at energies $E=$(1-10) TeV might distinguish between point-like and extended $\gamma$-ray emission. In case of extended emission, the determination of the centroid will allow to determine if it originates from the nucleus or from the more extended starburst ring.\\
Another  way to directly test  the   AGN  outflow model  is to look for neutrino signal. 
The fluxes predicted by this model  indicate that the neutrino signal from  \textsc{NGC 1068}  may be  detectable by the next generation neutrino telescope  KM3NeT, which, thanks to the improved angular resolution  compared to the current neutrino detectors, will allow to constrain effectively the possible astrophysical sources of high energy neutrino events.

The AGN outflow model can also be indirectly tested. A potential test is to determine the CR ionization rate of the molecular medium in the CND.  In fact, an enhanced molecular ionization  of  the CND gas could  be an indication that the $\gamma$-ray emission has an hadronic component. Molecular line surveys toward the nucleus of \textsc{NGC 1068} at ALMA resolution are therefore  necessary to properly determine the chemical and physical properties of the CND gas.\\
Finally,  the  large efficiencies required to accelerate protons and electrons in the  AGN outflow model  imply a large production of CR particles. An immediate consequence is that weaker magnetic fields are  required to produce the observed synchrotron emission in the radio continuum.  \\

\section*{Acknowledgements} 
We thank the Referee for his/her valuable comments that helped to improve the manuscript. A.L. and F.F. thank G.C. Perola for enlightening discussions. This work was supported by PRIN INAF 2014.
S.C. acknowledges support by the South African Research Chairs Initiative
of the Department of Science and Technology and National Research
Foundation and by the Square Kilometre Array (SKA).
This work is based on the research supported by the South African Research
Chairs Initiative of the Department of Science and Technology and National
Research Foundation of South Africa (Grant No 77948).
\bibliographystyle{aa}
\bibliography{biblio.bib}

\end{document}